\documentclass[10pt]{article}
\usepackage{latexsym}
\usepackage{hyperref}

\newtheorem{Algorithm}{Algorithm}[section]

\newtheorem{Theorem}[Algorithm]{Theorem}

\newtheorem{Proposition}[Algorithm]{Proposition}
\newtheorem{Example}[Algorithm]{Example}
\newenvironment{proof}{{\em Proof. }}{\hfill $\Box$ \par \medskip }
\newcommand\pd[2]{\frac{\partial#1}{\partial#2}}
\begin{document}

\title{Normal and seminormal forms of $sl_3$-valued zero curvature representations}
\author{Peter Sebesty\'en}
\date{\normalsize Mathematical Institute of the Silesian University in Opava, 
Czech Republic 
\\ e-mail: {\tt Peter.Sebestyen@math.slu.cz}}
\maketitle
\begin{abstract} 
We find normal and seminormal forms for a $sl_3$-valued zero curvature representation (ZCR).
We prove a theorem about reducibility of ZCR's, which says that if one of the matrix in a 
ZCR $(A,B)$ falls to a proper subalgebra of $sl_3$, then the second matrix either falls to 
the same subalgebra or the ZCR is almost trivial. In the end of this paper we show examples 
of ZCR's and their normal forms.
\end{abstract}

\section{Introduction}

Zero curvature representations (ZCR) rank among the most important 
attributes of integrable partial differential equations~\cite{T-F}.
A ZCR is usually treated as a special case of the Wahlquist--Estabrook 
prolongation structure~\cite{W-E}, but the famous Wahlquist--Estabrook 
procedure is not sufficient for obtaining a complete classification 
of integrable systems.
The main obstacle consists in the presence of a large group of gauge
transformations.
Thus we are naturally led to the problem of introduction of normal forms
of ZCR's such that every orbit of the gauge action contains the corresponding
normal form.

In nineties, independently M. Marvan~\cite{M93} and S. Yu. Sakovich~\cite{Sak}
introduced a characteristic element of 
a ZCR, which is a matrix that transforms by conjugation during gauge 
transformations of the ZCR. 
It follows that one can reduce the gauge freedom by putting 
the characteristic element in the Jordan normal form.
There is a remaining gauge freedom, which can be used for further
reduction of one of the matrices constituting the ZCR.
This is rather similar to classification of pairs of matrices under
simultaneous conjugation, developed by Belitski\u{\i}~\cite{Be}.

In case of the Lie algebra $sl_2$ a solution of the problem can be
found in \cite{M98}.
This made possible the subsequent complete classification of second-order
evolution equations possessing an $sl_2$-valued ZCR~\cite{M02}. 

In this work we try to obtain such a classification in case of $sl_3$.
The number of possible normal forms is 8, compared to 2 in case of $sl_2$.
As examples, we consider 
the Tzitz\'eica equation~\cite{Tz}, whose ZCR is known since 1910, 
Sawada-Kotera equation~\cite{Sats} and the Kupershmidt equation~\cite{For}.

\section{Preliminaries}

Let us consider a system of nonlinear differential equations
\begin{equation}
F^{l}(t,x,u^k,\dots,u_I^k,\dots)=0,   \label{eq1}
\end{equation}
in two independent variables $t$ and $x$, a finite number of dependent variables $u^k$ and
their derivatives $u_I^k$, where $I$ denotes a finite symmetric multiindex over $t$ 
and~$x$.

Let $J^{\infty}$ be an infinite-dimensional jet space such that $t$, $x$, $u^k$, $u_I^k$ are
local jet coordinates on $J^{\infty}$. We have two distinguished vector fields on $J^{\infty}$
$$D_t=\pd{}{t}+\sum_{k,I} u_{It}^k\pd{}{u_I^k} ,\qquad
D_x=\pd{}{x}+\sum_{k,I} u_{Ix}^k\pd{}{u_I^k},$$
which are called {\it total derivatives.}
Let $g$ be a matrix Lie algebra. By a $g$-valued {\it zero curvature representation} (ZCR)
for (\ref{eq1}) we mean two $g$-valued functions $A,B$ which satisfy
\begin{equation}
D_tA-D_xB+[A,B]=0     \label{eq2}
\end{equation}
as a consequence of (\ref{eq1}).
Let $G$ be the connected and simply connected matrix Lie group associated with $g$. Then
for every $G$-valued function $W$ we define the {\it gauge transformation} of ZCR $(A,B)$ by the formulas
\begin{eqnarray*}
A^W &:=& D_xW \cdot W^{-1} + W \cdot A \cdot W^{-1} \\
B^W &:=& D_tW \cdot W^{-1} + W \cdot B \cdot W^{-1} 
\end{eqnarray*}
As is well known, $(A^W,B^W)$ is a ZCR again, and we say that it is {\it gauge equivalent} to $(A,B).$ 

A {\it characteristic element} $R$ is a $g$-valued function defined in \cite{M93}. The following assertion holds:

\begin{Proposition} {\rm(\cite{M93})} 
Gauge equivalent ZCR's have conjugate characteristic elements.
\end{Proposition}

If a ZCR $(A,B)$ is gauge equivalent to another ZCR with coefficients in a proper subalgebra
of $g$, then we say that the ZCR is {\it reducible.} Otherwise it is said to be {\it
irreducible.} A ZCR gauge equivalent to zero is called {\it trivial.} A very important
case is a ZCR with coefficients in a non-solvable Lie algebra. The simplest case of a
non-solvable Lie algebra is the algebra $sl_2.$ In \cite{M98} the following proposition was obtained:

\begin{Proposition} Let $(A,B)$ be an irreducible $sl_2$-valued ZCR, let $R \neq 0$ be its
characteristic element. Then we have one of the two following normal forms for $R$ and
$A:$\\-- Nilpotent case
$$R=\left(\begin{array}{@{}cc@{}} 0&0 \\ 1&0 \end{array}\right),\qquad A=\left(\begin{array}{@{}cc@{}} 0&a_2 \\ a_3&0
\end{array}\right).$$
-- Diagonal case
$$R=\left(\begin{array}{@{}cc@{}} r&0 \\ 0&-r \end{array}\right),\qquad A=\left(\begin{array}{@{}cc@{}} a_1&1 \\ a_3&-a_1
\end{array}\right).$$
\end{Proposition}

\section{Basic notions}

In this section we explain the method to find the normal form of $g$-valued ZCR.
The main idea is taken from the proposition 2.1. Gauge
equivalent ZCR's have conjugate characteristic elements, therefore we can restrict ourselves to
the characteristic elements in the Jordan normal form. Since the gauge transformation is a
group action, it is possible to consider the stabilizer group of the characteristic element,
which is a proper subgroup of $G.$ 
The stabilizer is usually rather small (see Table 1), therefore we can
compute its action on the matrix $A$ and find the corresponding normal forms rather easily. 
We aim at finding the minimal set of normal forms.
We can achieve substantial reduction by taking
into account permutations of the  Jordan blocks and using suitable automorphism of $sl_3$.
$$
\begin{array}{l@{\hspace{1cm}}l}
J_1=\left(\begin{array}{@{}ccc@{}}\lambda_1&0&0 \\ 0&\lambda_2&0 \\ 0&0&-\lambda_1-\lambda_2 \end{array}\right);\quad
\lambda_1\neq\lambda_2,& 
W_1=\left(\begin{array}{@{}ccc@{}} w_1 &0&0 \\0& w_2 &0 \\ 0&0& (w_1w_2)^{-1} \end{array}\right), \\\\
J_2=\left(\begin{array}{@{}ccc@{}}\lambda&0&0 \\ 0&\lambda&0 \\ 0&0&-2\lambda \end{array}\right);\quad \lambda\neq 0 , & 
W_2=\left(\begin{array}{@{}ccc@{}} w_{11} & w_{12} &0 \\ w_{21} & w_{22} &0 \\ 0&0& Z^{-1} \end{array}\right), \\\\
J_3=\left(\begin{array}{@{}ccc@{}}\lambda&0&0 \\ 1&\lambda&0 \\ 0&0&-2\lambda \end{array}\right);\quad \lambda\neq 0 , & 
W_3=\left(\begin{array}{@{}ccc@{}} w_1 &0&0 \\ w_2 & w_1 &0 \\ 0&0& w_1^{-2} \end{array}\right), \\\\
J_4=\left(\begin{array}{@{}ccc@{}}0&0&0 \\ 1&0&0 \\ 0&0&0 \end{array}\right),& 
W_4=\left(\begin{array}{@{}ccc@{}} w_1 &0&0 \\ w_2 & w_1 & w_3 \\ w_4 &0& w_1^{-2} \end{array}\right), \\\\
J_5=\left(\begin{array}{@{}ccc@{}}0&0&0 \\ 1&0&0 \\ 0&1&0 \end{array}\right),& 
W_5=\left(\begin{array}{@{}ccc@{}} 1 &0&0 \\ w_2 & 1 &0 \\ w_3 & w_2 & 1 \end{array}\right), \\\\
{\rm where} \quad Z = w_{11}w_{22}-w_{12}w_{21}. & 
\end{array}
$$
\centerline{Table 1: Jordan forms and the corresponding stabilizers}

In this work we distinguish between {\it normal forms} and {\it seminormal forms}. We say that we have the
normal form if we have just a finite number of possibilities of the choice of the corresponding gauge matrix. If our choice
of the corresponding gauge matrix depends on at least one arbitrary function, we say that we have the seminormal form. In
this case we may use the residual gauge freedom to transform the matrix $B$.

Table 1 lists all possible Jordan forms $J_i$ of $sl_3$-matrices and the corresponding 
stabilizers $W_i$, where $w_j$ denote arbitrary complex numbers such that all algebraic operations
make sense.  $J_2$ and  $J_4$ are degenerate cases of $J_1$ and  $J_3$, respectively, when the two
eigenvalues coincide and  the dimension of the stabilizer raises from two to four.

\section{Reducibility theorem}

For further reference, we list here several subalgebras of $sl_3$.
Two subalgebras $a,b$ are said to be {\it conjugate}, if there exist $S \in GL_3$ such that $a= SbS^{-1}$.
Note that for constant matrices $S \in SL_3$ conjugation and gauge equivalence coincide.
Another obvious automorphism of $sl_3$ is $A \mapsto -A^\top$, which we call {\it transposition}.
We introduce six permutation matrices

$$\begin{array}{lll}
P_{0}=\left(\begin{array}{@{}ccc@{}}1&0&0 \\ 0&1&0 \\ 0&0&1 \end{array}\right),&
P_{1}=\left(\begin{array}{@{}ccc@{}} 1 & 0 & 0 \\ 0 & 0 & 1 \\ 0 & 1 & 0 
\end{array}\right),&
P_{2}=\left(\begin{array}{@{}ccc@{}} 0 & 1 & 0 \\ 0 & 0 & 1 \\ 1 & 0 & 0 
\end{array}\right),\\\\
P_{3}=\left(\begin{array}{@{}ccc@{}} 0 & 1 & 0 \\ 1 & 0 & 0 \\ 0 & 0 & 1 
\end{array}\right),&
P_{4}=\left(\begin{array}{@{}ccc@{}} 0 & 0 & 1 \\ 1 & 0 & 0 \\ 0 & 1 & 0 
\end{array}\right),&
P_{5}=\left(\begin{array}{@{}ccc@{}} 0 & 0 & 1 \\ 0 & 1 & 0 \\ 1 & 0 & 0 
\end{array}\right).
\end{array}$$
In fact the permutation matrices $P_{1},P_{3},P_{5}$ have the determinant equal to $-1$, but this 
makes no harm since $-P_{1},-P_{3},-P_{5} \in SL_3$ and $A^{P_{i}}=A^{-P_{i}}$.

\medskip

In what follows, we frequently use six 6-dimensional subalgebras consisting of traceless matrices
of either of the forms:

$$
\begin{array}{lll}
L_1=\left(\begin{array}{@{}ccc@{}} . & 0 & 0 \\ . &.& . \\ . & . &. \end{array}\right),&
L_2=\left(\begin{array}{@{}ccc@{}} . & . & . \\ 0 &.& 0 \\ . & . &. \end{array}\right),&
L_3=\left(\begin{array}{@{}ccc@{}} . & . & . \\ . &.& . \\ 0 & 0 &. \end{array}\right),\\\\
L_4=\left(\begin{array}{@{}ccc@{}} . & . & . \\ 0 &.& . \\ 0 & . &. \end{array}\right),&
L_5=\left(\begin{array}{@{}ccc@{}} . & 0 & . \\ . &.& . \\ . & 0 &. \end{array}\right),&
L_6=\left(\begin{array}{@{}ccc@{}} . & . & 0 \\ . &.& 0 \\ . & . &. \end{array}\right).
\end{array}
$$
These six subalgebras are mutually isomorphic via transposition or conjugation.

\begin{Theorem} \label{th1}
If the matrix $A$ in the $sl_3$-valued ZCR $(A,B)$ belongs to one of subalgebras $L_i, i=1,\dots,6,$
then the ZCR is either reducible or is gauge equivalent to one with $A=0, D_x B=0$.
\end{Theorem}

\begin{proof}
Recall that the ZCR $(A,B)$ is reducible if both $A,B$ fall to the same proper
subalgebra or is gauge equivalent to such.
Let the matrix $A$ belongs to the subalgebra $L_1=\{a_{12}=0,a_{13}=0\}$.

If $b_{13}=0$ then from (\ref{eq2}) we have $a_{23}b_{12}=0$. If $b_{12}=0$ then we are done. 
If $b_{12} \neq 0$ then $a_{23}=0$.
Now, if $b_{23} = 0$, then $A$ falls to the same subalgebra $L_{6}$ as $B$ and we are done.
If $b_{23} \neq 0$ then we express from (\ref{eq2}) stepwise all remaining elements of the matrix $A$:
\begin{eqnarray*}
a_{11} &=& (D_x b_{12} + (b_{12} D_x b_{23} - b_{23} D_x b_{12})/3 b_{23})/b_{12},
\\
a_{22} &=& (b_{12} D_x b_{23} - b_{23} D_x b_{12})/(3 b_{23} b_{12}),
\\
a_{21} &=&   (b_{23} b_{12}^2 D_{tx}b_{23} + 2 b_{23}^2 b_{12} D_{tx}b_{12} - b_{12}^2 D_x b_{23} D_t b_{23} - 
2 b_{23}^2 D_x b_{12} D_t b_{12} \\&&
 - 3 b_{23}^2 b_{12}^2 D_x b_{11})/(3 b_{12}^3 b_{23}^2), 
\\
a_{32} &=&  
(2 b_{23} b_{12}^2 D_{tx}b_{23} + b_{23}^2 b_{12} D_{tx}b_{12} - 2 b_{12}^2 D_x b_{23} D_t b_{23} - 
b_{23}^2 D_x b_{12} D_t b_{12} \\&&
- 3 b_{23}^2 b_{12}^2 D_x b_{11} - 3 b_{23}^2 b_{12}^2 D_x b_{22})/(3 b_{23}^3 b_{12}^2), 
\\
a_{31} &=&  
(b_{23}^2 b_{12}^3 D_{ttx} b_{23} + 2 b_{23}^3 b_{12}^2 D_{ttx} b_{12} - 2 b_{23} b_{12}^3 D_t b_{23} D_{tx} b_{23} \\&&
+ b_{23}^2 b_{11} b_{12}^3 D_{tx} b_{23} - b_{23}^2 b_{12}^2 D_t b_{12} D_{tx} b_{23} - b_{23}^2 b_{12}^3 b_{22} D_{tx} b_{23}\\&&
- b_{23} b_{12}^3 D_x b_{23} D_{tt} b_{23} - 3 b_{23}^3 b_{12}^3 D_{tx} b_{11} - 2 b_{23}^3 b_{12}^2 b_{22} D_{tx} b_{12} \\&&
- 6 b_{23}^3 b_{12} D_t b_{12} D_{tx} b_{12} + 2 b_{23}^3 b_{11} b_{12}^2 D_{tx} b_{12} - 2 b_{23}^3 b_{12} D_x b_{12} D_{tt} b_{12} \\&&
+ 2 b_{12}^3 D_x b_{23} (D_t b_{23})^2 + b_{23} b_{12}^2 D_x b_{23} D_t b_{23} D_t b_{12} + 6 b_{23}^3 D_x
b_{12} (D_t b_{12})^2\\&&
- b_{23} b_{11} b_{12}^3 D_x b_{23} D_t b_{23} + b_{23} b_{12}^3 b_{22} D_x b_{23} D_t b_{23} + 3
b_{23}^3 b_{12}^2 D_x b_{11} D_t b_{12}\\&&
 - 2 b_{23}^3 b_{11} b_{12} 
D_x b_{12} D_t b_{12} + 2 b_{23}^3 b_{12} b_{22} D_x b_{12} D_t b_{12} - 3 b_{23}^3 b_{12}^4 D_x b_{21} \\&&
- 3 b_{23}^3 b_{11} b_{12}^3 D_x b_{11} + 3 b_{23}^3 b_{12}^3 
b_{22} D_x b_{11} - 3 b_{21} b_{23}^3 b_{12}^3 D_x b_{12})/(3 b_{23}^4 b_{12}^4). 
\end{eqnarray*}
The gauge matrix which sends $A$ to zero is then
$$W = \left(\begin{array}{@{}ccc@{}} 
b_{23}^{-1/3}b_{12}^{-2/3} & 0 & 0 \\ 
w_{21} & b_{12}^{1/3}b_{23}^{-1/3} & 0 \\ 
w_{31} &  w_{32} & 
b_{12}^{1/3}b_{23}^{2/3}
\end{array}\right),$$
where
\begin{eqnarray*}
w_{21} &=& \textstyle(-\frac{1}{3} D_t b_{23}/b_{23} - \frac{2}{3} D_t b_{12}/b_{12} + b_{11})/
(b_{12}^{\frac{2}{3}} b_{23}^{\frac{1}{3}}), 
\\
w_{31} &=& \textstyle(-\frac{1}{3} D_{tt}b_{23}/b_{23} - \frac{2}{3} D_{tt}b_{12}/b_{12} + (\frac{2}{3} D_t b_{23}/b_{23})^2 
+ (\frac{10}{9} D_t b_{12}/b_{12})^2 \\&&
\textstyle + \frac{4}{9} D_t b_{23} D_t b_{12}/(b_{23} b_{12}) -
\frac{2}{3} b_{11} D_t b_{23}/b_{23} + D_t b_{11} \\&&
\textstyle - \frac{4}{3} b_{11} D_t b_{12}/b_{12} + b_{21} b_{12} + b_{11}^2) / (b_{12}^{\frac{2}{3}} b_{23}^{\frac{1}{3}}),
\\
w_{32} &=& \textstyle(-\frac{2}{3} D_t b_{23}/b_{23} - \frac{1}{3} D_t b_{12}/b_{12} + b_{11} + b_{22})
b_{12}^{\frac{1}{3}}/b_{23}^{\frac{1}{3}}.
\end{eqnarray*}
Then $D_x B=0$ by equation (\ref{eq2}).

\medskip
Finally, if $b_{13} \neq 0$, then we apply the gauge transformation to the ZCR $(A,B)$ with gauge matrix
$$
W = \left(\begin{array}{@{}ccc@{}} 
1 & 0 & 0 \\ 
0 & b_{12}/b_{13} & 1 \\ 
0 &  1 & 0 
\end{array}\right),
$$
which keeps $a_{12}=0,a_{13}=0$, while $b_{13}$ will become zero.
\end{proof}

\section{Normal and seminormal forms}

We list here normal forms of $sl_3$-valued ZCR's with characteristic element in Jordan normal form in
either of the forms $J_1,\dots,J_5$. Matrices $N_i^j$ denote normal forms, where the lower indices
corresponds with Jordan normal forms. Dots denote arbitrary elements.

$$
\begin{array}{l@{\hspace{2cm}}@{}cc@{}}
Case \enspace J_1 &
N_1^1 = \left(\begin{array}{@{}ccc@{}} . & . & . \\ 1& . & . \\ . & 1 & . \end{array}\right) & \\\\
Case \enspace J_2 &
N_2^1 = \left(\begin{array}{@{}ccc@{}} 0 & 1 & 0 \\ . & . & 1 \\ . & . & .  \end{array}\right) & \\\\
Case \enspace J_3 &
N_3^1 = \left(\begin{array}{@{}ccc@{}} . & . & 1 \\ . & . & 0 \\ . & . & .  \end{array}\right),&
N_3^2 = \left(\begin{array}{@{}ccc@{}} 0 & . & 0 \\ . & . & 1 \\ . & 0 & .  \end{array}\right) \\\\
Case \enspace J_4 &
N_4^1 = \left(\begin{array}{@{}ccc@{}} 0 & . & 0 \\ . & . & 1 \\ . & 0 & .  \end{array}\right),&
N_4^3 = \left(\begin{array}{@{}ccc@{}} 0 & 0 & 1 \\ . & 0 & 0 \\ . & . & 0  \end{array}\right) \\\\
Case \enspace J_5 &
N_5^1 = \left(\begin{array}{@{}ccc@{}} 0 & 0 & . \\ . & . & . \\ . & . & .  \end{array}\right),&
N_5^2 = \left(\begin{array}{@{}ccc@{}} 0 & . & 0 \\ . & . & . \\ . & 0 & .  \end{array}\right)
\end{array}
$$

\begin{Theorem} \label{th2}
In a $sl_3$-valued ZCR such that its characteristic element has either of the Jordan normal forms $J_1,\dots,J_5$,
the matrix $A$ has one of the above normal forms $N_1^1, N_2^1, N_3^1, N_3^2, N_4^1, N_4^3, N_5^1, N_5^2,$ 
or $A$ satisfies assumptions of Theorem \ref{th1}.
\end{Theorem}

The remaining part of this paper is devoted to the proof of Theorem \ref{th2}.
In fact, we give an algorithm which assigns a normal or seminormal form to the matrix $A$.
This algorithm proves Theorem \ref{th2}. The symbol $W_i^j$ denotes the corresponding gauge matrix which 
sends the matrix $A$ to the normal (resp. seminormal) form $N_i^j$ (resp. $S_i^j$).

\subsection{Case $J_1$}

The diagonal Jordan normal form is unique up to the order of the elements on the
diagonal, i.e., up to conjugation with respect to one of the permutation matrices $P_0,\dots,P_5$. 
Given a matrix $A$, the corresponding gauge equivalent matrices will be $A_{i} =
D_{x}P_{i}. P_{i}^{-1}+P_{i}AP_{i}^{-1} = P_{i}AP_{i}^{-1} $, $i=0,1,\dots,5$, namely
$$\begin{array}{lll}
A_{0}=\left(\begin{array}{@{}ccc@{}} a_{11} & a_{12} & a_{13} \\ a_{21} &  a_{22} & a_{23} \\
a_{31} & a_{32} & a_{33} \end{array}\right),&
A_{1}=\left(\begin{array}{@{}ccc@{}} a_{11} & a_{13} & a_{12} \\ a_{31} &  a_{33} & a_{32} \\
a_{21} & a_{23} & a_{22} \end{array}\right),&
A_{2}=\left(\begin{array}{@{}ccc@{}} a_{22} & a_{23} & a_{21} \\ a_{32} &  a_{33} & a_{31} \\
a_{12} & a_{13} & a_{11} \end{array}\right),\\\\
A_{3}=\left(\begin{array}{@{}ccc@{}} a_{22} & a_{21} & a_{23} \\ a_{12} &  a_{11} & a_{13} \\
a_{32} & a_{31} & a_{33} \end{array}\right),&
A_{4}=\left(\begin{array}{@{}ccc@{}} a_{33} & a_{31} & a_{32} \\ a_{13} &  a_{11} & a_{12} \\
a_{23} & a_{21} & a_{22} \end{array}\right),&
A_{5}=\left(\begin{array}{@{}ccc@{}} a_{33} & a_{32} & a_{31} \\ a_{23} &  a_{22} & a_{21} \\
a_{13} & a_{12} & a_{11} \end{array}\right).
\end{array}$$
where $a_{33} = -a_{11}-a_{22}$ \quad (since $A$ is an $sl_3$ matrix).

{\it Case 1. } If there exists $i=0,1,\dots,5$ such that $a_{21} \neq 0$ and $a_{32} \neq 0$ in $A=A_{i}$, then
we have
$$N_1^{1}=\left(\begin{array}{@{}ccc@{}} . & . & . \\ 1& . & . \\ . & 1 & . \end{array}\right),
\quad 
W_1^{1}=\left(\begin{array}{@{}ccc@{}} 
a_{32}^{\frac{1}{3}}a_{21}^{\frac{2}{3}} & 0 & 0 \\
0 & a_{32}^{\frac{1}{3}}a_{21}^{\frac{-1}{3}} & 0 \\
0 & 0 & a_{32}^{\frac{2}{3}}a_{21}^{\frac{-1}{3}} \end{array}\right).$$
One easily sees that the matrix $W_1^{1}$ is unique up to the choice of cubic roots, hence $N_1^{1}$ is a
{\it normal form}.

{\it Case 2. } Otherwise, if there exists $i=0,1,\dots,5$ such that $a_{21} \neq 0$, $a_{32}=0$ and $a_{31}
\neq 0$ in $A=A_{i}$, then 
$$N_1^{2}=\left(\begin{array}{@{}ccc@{}}
. & 0 & 0 \\ 1 & . & 0 \\ 1 & 0 & . \end{array}\right),
\quad
W_1^{2}=\left(\begin{array}{@{}ccc@{}} a_{31}^{\frac{1}{3}} a_{21}^{\frac{1}{3}} &  0 & 0 \\
0 &a_{31}^{\frac{1}{3}} a_{21}^{\frac{-2}{3}} & 0 \\
0 &  0 & a_{21}^{\frac{1}{3}} a_{31}^{\frac{-2}{3}}
\end{array}\right).$$
We have used the fact that $a_{12} = 0, a_{13} = 0$ and $a_{23} = 0$ as well. Indeed, if  $a_{12} \neq 0$ (resp. $a_{13}
\neq 0$, resp. $a_{23} \neq 0$) in $A$, then, using the permutation matrix $P_{3}$ (resp. $P_{4}$, resp. $P_{1}$), we
would obtain the first case.
$N_1^{2}$ is a normal form and belongs to the intersection of subalgebras $L_5$ and $L_6$.

The case when $a_{21} \neq 0$, $a_{32}=0$, $a_{31}=0$ and $a_{23} \neq 0$ in some $A=A_{i}$ can be converted to Case 2
by using conjugation by $P_{3}$ after transposition $A \mapsto -A^\top$.

{\it Case 3. }
Otherwise, if there exists $i=0,1,\dots,5$ such that $a_{21} \neq 0$, $a_{32}=0$, $a_{31}=0$ and $a_{23}=0$ 
in $A=A_{i}$, then
$$S_1^{3}=\left(\begin{array}{@{}ccc@{}} . & . & 0 \\ 1 &.& 0 \\
0&0&. \end{array}\right).
\quad
W_1^3=\left(\begin{array}{@{}ccc@{}} a_{21} & 0 & 0 \\ 0 & 1 & 0 \\ 0 & 0 &  a_{21}^{-1}
\end{array}\right).$$
Indeed, using the same argument as in Case 2 we may assume that $a_{13}=0$.
However, the most general gauge matrix is
$$\left(\begin{array}{@{}ccc@{}} a_{21}w_2 & 0 & 0 \\ 0 & w_2 & 0 \\ 0 & 0 &  a_{21}^{-1}w_2^{-2}
\end{array}\right)$$
and depends on the choice of one arbitrary function $w_2$.
If we set $w_2 = 1$, then we obtain $W_1^{3}$. Hence $S_1^{3}$ is a {\it seminormal form.}
The matrix $S_1^3$ belongs to the intersection of subalgebras $L_3$ and $L_6$.

{\it Case 4. } If $a_{21}=0$ for all $A_{i}$, then all the off-diagonal elements must
be zero, therefore the seminormal form is
$$A=S_1^{4}=\left(\begin{array}{@{}ccc@{}} . & 0 & 0 \\ 0 &.& 0 \\ 0&0&. \end{array}\right).$$
The matrix $S_1^4$ belongs to the intersection of subalgebras $L_1, L_2$ and $L_3$.

\subsection{Case $J_2$}

The following case $J_2$ of Jordan normal form of characteristic element is singular and the number of parameters
in the corresponding stabilizer subgroup increases from two to four (see Table 1). In this case we apply the
automorphism $A \mapsto -A^\top$ to reduce the number of normal and seminormal forms.

Let $K = a_{13} D_x a_{23} - a_{23} D_x a_{13} + a_{11} a_{13} a_{23} - a_{21} a_{13}^2 + a_{12} a_{23}^2 -
a_{22} a_{13} a_{23}$,
$L = a_{32} D_x a_{31} - a_{31} D_x a_{32} + a_{11} a_{32} a_{31} + a_{21} a_{32}^2 - a_{12} a_{31}^2 - 
a_{22} a_{32} a_{31}$, and \\
$R = a_{13} a_{31} + a_{23} a_{32}$.

{\it Case 1. } If $K \neq 0$, then the normal form is
$$N_2^1 = 
\left(\begin{array}{@{}ccc@{}} 0 & 1 & 0 \\ . & . & 1 \\ . & . & .  \end{array}\right).
$$
The corresponding gauge matrix $W_2^1$ is found to be
$$w_{11} = -{a_{23}}{K^{-2/3}},  \quad  w_{12} = {a_{13}}{K^{-2/3}}, $$
$$
w_{21} = \textstyle(\frac{2}{3} a_{23} K^{-1} D_x K - D_x a_{23} - a_{11} a_{23} + a_{13} a_{21}){K^{-2/3}}, $$
$$
w_{22} = \textstyle(-\frac{2}{3} a_{13} K^{-1} D_x K + D_x a_{13} - a_{12} a_{23} + a_{22} a_{13}){K^{-2/3}}.$$

\medskip
The case when $K = 0,L \neq 0$ can be reduced to Case 1.
Indeed, using the automorphism $A \mapsto -A^\top$ we have $K \mapsto -L$ and $L \mapsto -K$.
\medskip

{\it Case 2. } If $K = 0, L = 0, R \neq 0$, then
$$S_2^2 = 
\left(\begin{array}{@{}ccc@{}} . & 0 & 0 \\ 0 & . & . \\ 0 & 1 & .  \end{array}\right),
\quad
W_2^2 =
\left(\begin{array}{@{}ccc@{}} a_{23} & -a_{13} & 0 \\ 
a_{31} R^{-1/2} & a_{32} R^{-1/2} & 0 \\ 
0 & 0 & R^{-1/2}
 \end{array}\right).
$$
Indeed,  applying $W_{2}^2$ to general $sl_3$ matrix $A$ we obtain 
$$A^{W_2^{2}} = 
\left(\begin{array}{@{}ccc@{}}
 . & -K R^{-1/2} & 0 \\ L R^{-3/2} & . & . \\ 0 & 1 & .  
\end{array}\right),
$$
and we see that for $K = 0, L = 0$ we have $A^{W_{2}^2} = S_{2}^2$.
The seminormal form $S_2^2$ falls to the intersection of subalgebras $L_1$ and $L_4$.

For $K = 0, L = 0, R = 0$ we have two subcases:

{\it Case 3. } If $a_{13} \neq 0$ or $a_{23}\neq 0$, then
$$S_{2}^{3} = 
\left(\begin{array}{@{}ccc@{}} . & 0 & 0 \\ . & . & . \\ . & 0 & .  \end{array}\right),
\quad
W_2^{3} =
\left(\begin{array}{@{}ccc@{}} a_{23} & -a_{13} & 0 \\ 
{w_{21}} & {w_{22}} & 
0 \\ 0 & 0 & (w_{21} a_{13} + w_{22} a_{23})^{-1}
 \end{array}\right)
$$
for arbitrary nonzero parameters $w_{21}$ and $w_{22}$ such that $w_{21} a_{13} + w_{22} a_{23} \neq 0$. Indeed, applying
$W_{2}^{3}$ to general $sl_3$ matrix $A$ we obtain 
$$A^{W_2^{3}} = 
\left(\begin{array}{@{}ccc@{}} . & K(w_{21} a_{13} + w_{22} a_{23})^{-1} & 0 \\ 
. & . & . \\ 
. &  R (w_{21} a_{13} + w_{22} a_{23})^{-2} & . 
\end{array}\right),
$$
and we see that for $K = 0, R = 0$ we have $A^{W_{2}^{3}} = S_{2}^{3}$. Note that in this case 
$L = -K (a_{32} / a_{13})^2$ or $L = -K (a_{31} / a_{23})^2$.
The seminormal form $S_2^{3}$ falls to the intersection of subalgebras $L_1$ and $L_5$.

\medskip
When $a_{31} \neq 0$ or $a_{32}\neq 0$, then using the transposition $A \mapsto -A^\top$ we obtain Case 3.
\medskip

{\it Case 4. } If $a_{13} = 0, a_{23} = 0, a_{31} = 0, a_{32} = 0$, then the
seminormal form is
$$A=S_{2}^{4} = 
\left(\begin{array}{@{}ccc@{}} . & . & 0 \\ . & . & 0 \\ 0 & 0 & .  \end{array}\right),
$$
The seminormal form $S_2^{4}$ falls to the intersection of subalgebras $L_3$ and $L_6$.

\subsection{Case $J_3$} 

In this case we use a modification of the permutation matrix $P_{3}$:
$$\overline{P_{3}} = 
\left(\begin{array}{@{}ccc@{}} 0 & 1 & 0 \\ -1 & 0 & 0 \\ 0 & 0 & 1 
\end{array}\right).$$

{\it Case 1. } If $a_{13} \neq 0$, then
$$N_3^1 = 
\left(\begin{array}{@{}ccc@{}} . & . & 1 \\ . & . & 0 \\ . & . & .  \end{array}\right),
\quad
W_3^1 = \left(\begin{array}{@{}ccc@{}} a_{13}^{-1/3} & 0 & 0 \\ 
-a_{23} a_{13}^{-4/3} & a_{13}^{-1/3} & 0 \\ 
0 & 0 & a_{13}^{2/3} 
\end{array}\right).
$$
$N_3^1$ is a normal form.

The case when $a_{13} = 0 , a_{32} \neq 0$ can be reduced to Case 1 by using the conjugation by 
$\overline{P_{3}}$ after transposition $A \mapsto -A^\top$.

{\it Case 2. } If $a_{13} = 0 , a_{32} = 0 , a_{23} \neq 0 , a_{12} \neq 0$, then
$$N_3^2 = 
\left(\begin{array}{@{}ccc@{}} 0 & . & 0 \\ . & . & 1 \\ . & 0 & .  \end{array}\right),
\quad
W_3^2 =
\left(\begin{array}{@{}ccc@{}} a_{23}^{-1/3} & 0 & 0 \\ 
(a_{23} a_{11} - \frac{1}{3} D_x a_{23}) / (a_{12} a_{23}^{4/3}) & 
a_{23}^{-1/3} & 0 \\ 
0 & 0 & a_{23}^{2/3} 
\end{array}\right).
$$
$N_3^2$ is a normal form.

{\it Case 3. } If $a_{13} = 0 , a_{32} = 0 , a_{23} \neq 0 , a_{12} = 0$, then
$$S_3^3 = 
\left(\begin{array}{@{}ccc@{}} . & 0 & 0 \\ . & . & 1 \\ . & 0 & .  \end{array}\right),
\quad
W_3^3 =
\left(\begin{array}{@{}ccc@{}} a_{23}^{-1/3} & 0 & 0 \\ 
0 & a_{23}^{-1/3} & 0 
\\ 0 & 0 & a_{23}^{2/3} 
\end{array}\right).
$$
$S_3^3$ is a seminormal form and belongs to the intersection of subalgebras $L_1$ and $L_5$.

{\it Case 4. } If $a_{13} = 0 , a_{32} = 0 , a_{23} = 0 , a_{31} \neq 0 , a_{12} \neq
0$, then
$$N_3^4 = 
\left(\begin{array}{@{}ccc@{}} 0 & . & 0 \\ . & . & 0 \\ 1 & 0 & .  \end{array}\right),
\quad
W_3^4 =
\left(\begin{array}{@{}ccc@{}} a_{31}^{1/3} & 0 & 0 \\ 
(\frac{1}{3} D_x a_{31} + a_{31} a_{11})/(a_{12} a_{31}^{2/3}) & 
a_{31}^{1/3} & 0 \\ 0 & 0 & a_{31}^{-2/3} 
\end{array}\right).
$$
$N_3^4$ is a normal form and belongs to the subalgebra $L_6$.

The case when $a_{13} = 0 , a_{32} = 0 , a_{23} = 0 , a_{31} \neq 0 , a_{12} = 0$ can be reduced to Case 3
by using the conjugation by $\overline{P_{3}}$ after transposition $A \mapsto -A^\top$ again.

{\it Case 5. } Otherwise, if $a_{13} = 0 , a_{32} = 0 ,  a_{23} = 0 , a_{31} = 0 , a_{12} \neq 0$, then
$$S_3^5 = 
\left(\begin{array}{@{}ccc@{}} 0 & . & 0 \\ . & . & 0 \\ 0 & 0 & .  \end{array}\right),
\quad
W_3^5 =
\left(\begin{array}{@{}ccc@{}} 1 & 0 & 0 \\ a_{11} a_{12}^{-1} & 
1 & 0 \\ 0 & 0 & 1 \end{array}\right).
$$
$S_3^5$ is a seminormal form and belongs to the intersection of subalgebras $L_3$ and $L_6$.

{\it Case 6. } Otherwise, if $a_{13} = 0 , a_{32} = 0 , a_{23} = 0 , a_{31} = 0 , a_{12} = 0 $,
then the seminormal form is
$$A=S_3^6 = 
\left(\begin{array}{@{}ccc@{}} . & 0 & 0 \\ . & . & 0 \\ 0 & 0 & .  \end{array}\right).
$$
Matrices of this form constitute a 3-dimensional solvable subalgebra of $sl_3$.

\subsection{Case $J_4$}

The Case $J_4$ is singular again (see Case $J_2$). We use again the modified permutation matrix $\overline{P_{3}}$.

Let $M = a_{12} D_x a_{13} - a_{13} D_x a_{12} - 2 a_{12} a_{13} a_{22} + a_{23} a_{12}^2 - a_{32} a_{13}^2 -
a_{11} a_{12} a_{13}$ and
$N = a_{12} D_x a_{32} - a_{32} D_x a_{12} +  2 a_{11} a_{12} a_{32} - a_{31} a_{12}^2 + a_{13} a_{32}^2 +
a_{12} a_{22} a_{32}$.

{\it Case 1. } If $a_{12} \neq 0, M \neq 0$, then the normal form is
$$N_4^1 = 
\left(\begin{array}{@{}ccc@{}} 0 & . & 0 \\ . & . & 1 \\ . & 0 & .  \end{array}\right).
$$
The corresponding gauge matrix $W_4^1$ is obtained in the following way:
$$w_1 = a_{12}^{2/3}/ M^{1/3}, \qquad
w_2 = (D_x w_1 + a_{11} w_1)/a_{12}, $$
$$
w_3 = w_1 a_{13}/a_{12}, \qquad
w_4 = -a_{32}/(w_1^2 a_{12}).
$$

The case when $a_{12} \neq 0, M = 0, N \neq 0$ may be reduced to Case 1 by using the conjugation by
$\overline{P_{3}}$ after transposition $A \mapsto -A^\top$.

{\it Case 2. } If $a_{12} \neq 0, M = 0, N = 0$, then
$$S_4^2 = 
\left(\begin{array}{@{}ccc@{}} 0 & . & 0 \\ . & . & 0 \\ 0 & 0 & .  \end{array}\right),
\quad
W_4^2 = 
\left(\begin{array}{@{}ccc@{}} 1 & 0 & 0 \\ 
a_{11}a_{12}^{-1} & 1 & a_{13}a_{12}^{-1} \\ 
-a_{32}a_{12}^{-1} & 0 & 1 
\end{array}\right).
$$
Indeed,  applying $W_{4}^2$ to general $sl_3$ matrix $A$ we obtain 
$$A^{W_4^{2}} = 
\left(\begin{array}{@{}ccc@{}}
0 & . & 0 \\ . & . & M/ a_{12}^2 \\ -N/ a_{12}^2 & 0 & .  
\end{array}\right),
$$
and we see that for $M = 0, N = 0$ we have $A^{W_{4}^2} = S_{4}^2$.
The seminormal form $S_4^2$ falls to the intersection of subalgebras $L_3$ and $L_6$.

{\it Case 3. } If $a_{12} = 0, a_{13} \neq 0, a_{32} \neq 0$, then the normal form is
$$N_4^3 = 
\left(\begin{array}{@{}ccc@{}} 0 & 0 & 1 \\ . & 0 & 0 \\ . & . & 0  \end{array}\right).
$$
The corresponding gauge matrix $W_4^3$ is obtained in the following way:
$$w_1 = {a_{13}^{-1/3}}, \qquad  w_3 = (D_x a_{13} - 3 a_{13} a_{22})/3 a_{32} a_{13}^{4/3},$$
$$w_4 = -(D_x a_{13} - 3 a_{11} a_{13})/{3 a_{13}^{4/3}},$$
$$
w_2 = -\frac{w_3 D_x a_{13}}{3 a_{13}^2} - 
\frac {D_x w_3 - a_{32} a_{13}^{1/3} w_3^2 - a_{11} w_3 - 2 a_{22} w_3 + a_{23} a_{13}^{-1/3}}{a_{13}}.$$

{\it Case 4. } If $a_{12} = 0, a_{13} = 0, a_{32} \neq 0$, then
$$S_4^4 = 
\left(\begin{array}{@{}ccc@{}} . & 0 & 0 \\ . & . & . \\ 0 & 1 & 0  \end{array}\right),
\quad
W_4^4 = \left(\begin{array}{@{}ccc@{}} a_{32}^{1/3} & 0 & 0 \\ 
\displaystyle \frac{a_{31}}{a_{32}^{2/3}} & a_{32}^{1/3} & 
\displaystyle -\frac{2 D_x a_{32} + 3 a_{32} a_{11} + 3 a_{32} a_{22}}{3 a_{32}^{5/3}} \\ 
0 & 0 & a_{32}^{-2/3} \end{array}\right).
$$
The seminormal form $S_4^4$ falls to the subalgebra $L_1$.

The case when $a_{12} = 0, a_{13} \neq 0, a_{32} = 0$ is reducible to Case 4 by using the conjugation by
$\overline{P_{3}}$ after transposition $A \mapsto -A^\top$.

{\it Case 5. } If $a_{12} = 0, a_{13} = 0, a_{32} = 0, a_{23} \neq 0$, then
$$S_4^5 = 
\left(\begin{array}{@{}ccc@{}} . & 0 & 0 \\ 0 & . & . \\ . & 0 & .  \end{array}\right),
\quad
W_4^5 = \left(\begin{array}{@{}ccc@{}} 1 & 0 & 0 \\ 0 & 1 & 0 \\ 
a_{21} a_{23}^{-1} & 0 & 1 \end{array}\right).
$$
The seminormal form $S_4^5$ falls to the intersection of subalgebras $L_1$ and $L_5$.

{\it Case 6. } If $a_{12} = 0, a_{13} = 0, a_{32} = 0, a_{23} = 0$, then the seminormal form is
$$A=S_4^6 = 
\left(\begin{array}{@{}ccc@{}} . & 0 & 0 \\ . & . & 0 \\ . & 0 & .  \end{array}\right).
$$
The seminormal form $S_4^6$ falls to the intersection of subalgebras $L_5$ and $L_6$.

\subsection{Case $J_5$}

In this case we use a modification of the permutation matrix $P_{5}$:
$$\overline{P_{5}} = 
\left(\begin{array}{@{}ccc@{}} 0 & 0 & 1 \\ 0 & -1 & 0 \\ 1 & 0 & 0 
\end{array}\right).$$

{\it Case 1. } If $a_{13} \neq 0$, then
$$N_5^1 = 
\left(\begin{array}{@{}ccc@{}} 0 & 0 & . \\ . & . & . \\ . & . & .  \end{array}\right),
\quad
W_5^1 = 
\left(\begin{array}{@{}ccc@{}} 1 & 0 & 0 \\ a_{12} a_{13}^{-1} & 1 & 0 \\ 
a_{11} a_{13}^{-1} & a_{12} a_{13}^{-1} & 1 
\end{array}\right)
$$
$N_5^1$ is a normal form.

{\it Case 2. } Otherwise, if $a_{13} = 0 , a_{12} \neq 0$, then
$$N_5^2 = 
\left(\begin{array}{@{}ccc@{}} 0 & . & 0 \\ . & . & . \\ . & 0 & .  \end{array}\right),
\quad
W_5^2 =
\left(\begin{array}{@{}ccc@{}} 1 & 0 & 0 \\ a_{11} a_{12}^{-1} & 1 & 0 \\ 
w_3 & a_{11} a_{12}^{-1} & 1 \end{array}\right),
$$
where $w_3 = (a_{11} D_x a_{12} - a_{12} D_x a_{11} + a_{23} a_{11}^2 - a_{32} a_{12}^2 - 2 a_{22} a_{12} a_{11} - a_{12}
a_{11}^2)/{a_{12}^3}$.
$N_5^2$ is a normal form.

The case when $a_{13} = 0 , a_{12} = 0 , a_{23} \neq 0$ is reducible to Case 2 by using the conjugation by
$\overline{P_{5}}$ after transposition $A \mapsto -A^\top$.

{\it Case 3. } Otherwise, if $a_{13} = 0 , a_{12} = 0 , a_{23} = 0$, then the seminormal form is
$$A=S_5^3 = 
\left(\begin{array}{@{}ccc@{}} . & 0 & 0 \\ . & . & 0 \\ . & . & .  \end{array}\right).
$$
Matrices of this form fall to the intersection of subalgebras $L_1$ and $L_6$.

\section{Examples}

\begin{Example}
\rm The Tzitz\'eica equation~\cite{Tz}:
$$u_{tx}=e^u-e^{-2u}.$$ 
The corresponding ZCR, which depends on a parameter $m \ne 0$, is
$$A=\left(\begin{array}{@{}ccc@{}} -u_x & 0 & m \\ m & u_x & 0 \\ 0 & m & 0
\end{array}\right),\quad
B=\left(\begin{array}{@{}ccc@{}} 0 & {\mathrm e^{-2 u}}/m & 0 \\ 
0 & 0 & {\mathrm e^{u}}/m \\ {\mathrm e^{u}}/m & 0 & 0 
\end{array}\right). $$
The matrix $A$ belongs to the Case $J_1$ with the normal form $N_1^1.$
Namely, the Jordan normal form of the characteristic element $R$ and the matrix $A^{W_1^1}$ are
$$
R =\left(\begin{array}{@{}ccc@{}} -1 & 0 & 0 \\ 0 & 1 & 0 \\ 0 & 0 & 0
 \end{array}\right) ,\quad
A^{W_1^1} =\left(\begin{array}{@{}ccc@{}} -u_x & 0 & m^3 \\ 1 & u_x & 0 \\ 0 & 1 & 0
 \end{array}\right) .
$$
\end{Example}

\begin{Example}\rm The Sawada-Kotera equation~\cite{Sats}:
$$u_t = u_{xxxxx} + 5 u u_{xxx} + 5 u_x u_{xx}+5 u^2 u_x$$ 
The corresponding ZCR and the corresponding normal form belongs to the Case $J_4$. The matrix $A$ has the 
normal form $N_4^1,$ namely,
$$A=\left(\begin{array}{@{}ccc@{}} 0 & -1 & 0 \\ u & 0 & -m \\ 1 & 0 & 0 
\end{array}\right),\quad
A^{W_4^1} =\left(\begin{array}{@{}ccc@{}} 0 & -1 & 0 \\ u & 0 & 1 \\ -m & 0 & 0 
\end{array}\right).$$
The matrix $B$ is rather large, hence omitted.
\end{Example}

\begin{Example}
\rm The Kupershmidt equation~\cite{For}:
$$u_t = u_{xxxxx} + 10 u u_{xxx} + 25 u_x u_{xx} + 20 u^2 u_x$$ 
The corresponding ZCR and the corresponding normal form belongs to the Case $J_5$. The matrix $A$ has the normal form $N_5^2,$ namely,
$$A=\left(\begin{array}{@{}ccc@{}} 0 & 1 & 0 \\ -u & 0 & 1 \\ m & -u & 0 
\end{array}\right), \quad
A^{W_5^2} =\left(\begin{array}{@{}ccc@{}} 0 & 1 & 0 \\ -2 u & 0 & 1 \\ u_x + m & 
0 & 0 \end{array}\right).$$
The matrix $B$ is large, hence omitted again.
\end{Example}

\section*{Acknowledgements}

The support from the grant MSM:J10/98:192400002 is gratefully acknowledged.
This work was very much influenced by discussions with M. Marvan and A. Sergyeyev.


\begin{thebibliography}{99}



\bibitem{Be}
G.R. Belitski\u{\i}, Normal forms in a space of matrices, in:
{\it Analysis in Infinite-dimensional Spaces and Operator Theory}
(Naukova Dumka, Kiev, 1983) 3--15 (in Russian).

\bibitem{For}
A.P. Fordy and J. Gibbons,
Some remarkable nonlinear transformations, {\it Phys. Lett. A} {\bf 75} (1979/80), no. 5, 325. 

\bibitem{M93}
M. Marvan, 
On zero curvature representations of partial differential equations, 
in: {\it Differential Geometry and Its Applications} 
Proc. Conf. Opava, Czechoslovakia, Aug. 24--28  1992 
(Opava: Silesian University, 1993) 
103--122 (online ELibEMS http://www.emis.de/proceedings)

\bibitem{M98}
M. Marvan, 
A direct procedure to compute zero-curvature 
representations. The case sl$_2$, in: {\it Secondary Calculus
and Cohomological Physics}, Proc. Conf. Moscow 1997 
(online ELibEMS http://www.emis.de/proceedings/SCCP97), 1998, pp. 10.

\bibitem{M02}
M. Marvan, 
Scalar second order evolution equations possessing an irreducible 
sl$_2$-valued zero curvature representation, 
{\it J. Phys. A: Math. Gen.} {\bf 35} (2002) 9431--9439.

\bibitem{Sak}
S. Yu. Sakovich,
On zero-curvature representations of evolution equations,
{\it J. Phys. A: Math. Gen.} {\bf 28} (1995) 2861--2869.

\bibitem{Sats}
J. Satsuma and D.J. Kaup,
A B\"acklund transformation for a higher order Korteweg-de Vries equation,
{\it J. Phys. Soc. Japan} {\bf 43} (1977), no. 2, 692--726.

\bibitem{T-F}
L.A. Takhtadzhyan and L.D. Faddeev,
{\it Hamiltonian Methods in the Theory of Solitons} 
(Springer, Berlin et al., 1987).

\bibitem{Tz} 
G. Tzitz\'eica,
Sur une nouvelle classe de surface, {\it C.R. Acad. Sci. Paris}
{\bf 150} (1910) 955--956, 1227--1229.

\bibitem{W-E} 
H.D. Wahlquist and F.B. Estabrook,
Prolongation structures and nonlinear evolution equations I, II,
{\it J. Math. Phys.} {\bf 16} (1975) 1--7; {\bf 17} (1976) 1293--1297.

\end{thebibliography}
\end{document}